\documentclass[12pt,preprint]{aastex}
\usepackage{lineno}
\linenumbers
\shortauthors{The \emph{Fermi}/LAT Collaboration}

\begin{document}

 
\title{Radio-Loud Narrow-Line Seyfert 1\\as a New Class of Gamma-Ray AGN}


\author{
A.~A.~Abdo\altaffilmark{1,2}, 
M.~Ackermann\altaffilmark{3}, 
M.~Ajello\altaffilmark{3}, 
L.~Baldini\altaffilmark{4}, 
J.~Ballet\altaffilmark{5}, 
G.~Barbiellini\altaffilmark{6,7}, 
D.~Bastieri\altaffilmark{8,9}, 
K.~Bechtol\altaffilmark{3}, 
R.~Bellazzini\altaffilmark{4}, 
B.~Berenji\altaffilmark{3}
% ...
}
\altaffiltext{1}{Space Science Division, Naval Research Laboratory, Washington, DC 20375, USA}
\altaffiltext{2}{National Research Council Research Associate, National Academy of Sciences, Washington, DC 20001, USA}
\altaffiltext{3}{W. W. Hansen Experimental Physics Laboratory, Kavli Institute for Particle Astrophysics and Cosmology, Department of Physics and SLAC National Accelerator Laboratory, Stanford University, Stanford, CA 94305, USA}
\altaffiltext{4}{Istituto Nazionale di Fisica Nucleare, Sezione di Pisa, I-56127 Pisa, Italy}
\altaffiltext{5}{Laboratoire AIM, CEA-IRFU/CNRS/Universit\'e Paris Diderot, Service d'Astrophysique, CEA Saclay, 91191 Gif sur Yvette, France}
\altaffiltext{6}{Istituto Nazionale di Fisica Nucleare, Sezione di Trieste, I-34127 Trieste, Italy}
\altaffiltext{7}{Dipartimento di Fisica, Universit\`a di Trieste, I-34127 Trieste, Italy}
\altaffiltext{8}{Istituto Nazionale di Fisica Nucleare, Sezione di Padova, I-35131 Padova, Italy}
\altaffiltext{9}{Dipartimento di Fisica ``G. Galilei", Universit\`a di Padova, I-35131 Padova, Italy}
% ...

\begin{abstract}
We report the discovery with \emph{Fermi}/LAT of $\gamma-$ray emission from three radio-loud narrow-line Seyfert 1 galaxies: PKS~1502+036 ($z=0.409$), 1H~0323+342 ($z=0.061$) and PKS~2004-447 ($z=0.24$). In addition to PMN J0948+0022 ($z=0.585$), the first source of this type to be detected in $\gamma$ rays, they may form an emerging new class of $\gamma-$ray active galactic nuclei (AGN). These findings can have strong implications on our knowledge about relativistic jets and the unified model of AGN.
\end{abstract}


\section{Introduction}
The advent of the \emph{Fermi Gamma-ray Space Telescope} (hereafter, \emph{Fermi}), with its excellent performance, is changing our perception of the sky at high-energy $\gamma$ rays. Specifically, the recent detection in the MeV-GeV energy range of the radio-loud narrow-line Seyfert 1 (RL-NLS1) quasar PMN J0948+0022 strongly supports the presence of a closely aligned relativistic jet in this peculiar system (Abdo et al. 2009a,c; Foschini et al. 2009a). This is quite surprising, since NLS1s are generally hosted in spiral galaxies and the presence of a fully developed relativistic jet is contrary to the well-known paradigm that this type of system is associated to ellipticals (e.g. see Marscher 2009 for a recent review). 

Already in the past, several authors inferred from the multiwavelength properties of RL-NLS1s some parallelism between this type of source and blazars (e.g. Komossa et al. 2006, Yuan et al. 2008, Foschini et al. 2009b), suggesting similarities with different flavors of the blazar types (quasars or BL Lacs). What was missing in these previous studies is the $\gamma-$ray detection, which is important to confirm the presence of a relativistic jet, to measure its power, and to study the characteristics of this type of sources by modeling their spectral energy distributions (SEDs). Indeed, blazar-like radio emission from radio-quiet AGNs has been already reported (e.g., Brunthaler et al. 2000, Brunthaler et al. 2005, Lister et al. 2009; see also the review by Ho 2008), but no $\gamma$ rays have been yet detected from these sources. This suggests some evident differences between fully developed relativistic jets (i.e. emitting over the whole electromagnetic spectrum, from radio to $\gamma$ rays) and jet-like (perhaps aborted?, cf Ghisellini et al. 2004) structures, whose emission at $\gamma$ rays in the MeV-TeV energy range -- if any -- has not been yet reported. 

Understanding these differences could give important insights into jet formation. Therefore, we started a larger program aiming at the detection of $\gamma-$ray emission from other RL-NLS1s. Since no complete sample of RL-NLS1s is available in the literature, we have built a sample by merging all the sources of this type in the catalog of Zhou \& Wang (2002) and in the lists of Komossa et al. (2006) and Yuan et al. (2008). We have adopted a threshold in the radio-loudness $R_L=f_{\rm 5~GHz}/f_{\rm 440~nm}=50$, in order to avoid doubtful sources at the border ($R_L=10$) with radio-quietness. The optical properties are those typical of NLS1s, i.e. narrow permitted lines (FWHM H$\beta$$<2000$~km~s$^{-1}$), [OIII]/H$\beta <3$ and the bump of FeII (see Pogge 2000 for a review). At radio frequencies, RL-NLS1s display strong and variable radio emission, with brightness temperature well above the inverse-Compton limit (see, e.g., Yuan et al. 2008, Table 3). 

This selection resulted in a list of 29 sources, most of them selected from the \emph{Sloan Digital Sky Survey} (SDSS). This low number should not be a surprise. According to Komossa et al. (2006), RL-NLS1s are more rare than quasars: 7\% of NLS1s have $R_L>10$ and only 2.5\% exceed $R_L=50$, while generally 10-20\% of quasars are radio-loud. Although based on a small sample of objects (128), these results have been confirmed by studies on larger samples from the SDSS (Zhou et al. 2006, Whalen et al. 2006). It is worth noting that this sample is not complete, although the fact that it is based on the SDSS ensures a minimum degree of representativeness.

Here we report three new detections at $\gamma$ rays of RL-NLS1s obtained with \emph{Fermi}/LAT. In addition to the already known PMN J0948+0022 (reported by Abdo et al. 2009a,c; Foschini et al. 2009a), this increases the number of $\gamma-$ray detections of NLS1s ($3+1=4$ out of 29 sources, $\approx 14$\% of our sample), suggesting that they form a new class of $\gamma-$ray emitting AGNs. 

Throughout this work, we adopted a $\Lambda$CDM cosmology from the most recent \emph{WMAP} results, which give the following values for the cosmological parameters: $h = 0.71$, $\Omega_m = 0.27$, $\Omega_\Lambda = 0.73$ and with the Hubble-Lema\^{i}tre constant $H_0=100h$ km s$^{-1}$ Mpc$^{-1}$ (Komatsu et al. 2009).

\section{Data Analysis}
The data from the Large Area Telescope (LAT, Atwood et al. 2009) onboard \emph{Fermi} were analyzed following the same scheme described in Abdo et al. (2009a), but using a more recent version of the software ({\tt Science Tools v 9.15.2}), Instrument Response Function (\texttt{IRF P6\_V3\_DIFFUSE}, Rando et al. 2009) and background subtraction(\footnote{We adopted the same version of software and calibration database that is publicly available at \texttt{http://fermi.gsfc.nasa.gov/ssc/data/analysis/software/}.}). The data analyzed span between MJD 54682 (2008 August 4) and 55048 (2009 August 5). 

Four LAT sources with TS\footnote{Likelihood test statistic, see Mattox et al. (1996) for a definition of this test.} greater than $25$, which is equivalent\footnote{The significance of the detection in $\sigma$ is roughly $\sqrt{TS}$.} to $\sim 5\sigma$, can be associated with RL-NLS1s by using the Figure-of-Merit (FoM) method outlined in Abdo et al. (2009b) and already used to associate other LAT sources with radio counterparts. One of them is the already well-known PMN J0948+0022 (Abdo et al. 2009a). The three remaining new sources were associated to the RL-NLS1 objects PKS~1502+036 ($z=0.409$), 1H~0323+342 ($z=0.061$) and PKS~2004-447 ($z=0.24$). The results are summarized in Table~\ref{tab:latsources}.

\begin{table*}[!ht]
\caption{$\gamma-$ray characteristics of RL-NLS1s detected by LAT with $TS > 25$.\label{tab:latsources}}
\vskip 0.4 true cm
\centering
\scriptsize
\begin{tabular}{lccccccccc}
\hline
\hline
Name & $z$ & $\alpha$ & $\delta$ & $r_{\rm 95\%}$\tablenotemark{a} & $F_{\gamma}$\tablenotemark{b} & $\Gamma$\tablenotemark{c} & $TS$ & $L_{\gamma}$\tablenotemark{d} & Ass.\tablenotemark{e}\\
{}   & {}  & [deg] & [deg] & [deg]  & [$10^{-8}$~ph~cm$^{-2}$~s$^{-1}$] & {} & {} & [$10^{46}$~erg~s$^{-1}$] & [\%]\\   
\hline   
1H 0323+342 & 0.061 & $51.27$ & $+34.19$ & $0.21$ & $5.7\pm 0.3$ & $2.74\pm 0.03$ & 78 & 0.02 & 92\\
PKS 1502+036 & 0.409 & $226.26$ & $+3.37$ & $0.18$ & $7.5\pm 0.3$ & $2.81\pm 0.03$ & 186 & 2.1 & 93\\
PKS 2004-447 & 0.24 & $301.99$ & $-44.51$ & $0.09$ & $2.3\pm 0.3$ & $2.5\pm 0.4$ & 43 & 0.2 & 81\\
\hline  
PMN J0948+0022\tablenotemark{f} & 0.585 & $147.28$ & $+0.39$ & $0.12$ & $14.6\pm 0.9$ & $2.78\pm 0.06$ & 647 & 11 & 99\\
\hline
\end{tabular}
\tablenotetext{a}{Error radius at 95\% confidence level. To take properly into account systematic effects in the calculation of the error radii, we added the absolute systematic error of $0^{\circ}.0075$ in quadrature to the 68\%, then we convert to 95\% by multiplying by 1.62 and, finally, we multiply by 1.2, which is the relative systematic error (Abdo et al., in preparation).}
\tablenotetext{b}{$\gamma-$ray flux for $E>100$~MeV. We quoted only statistical errors, while the systematics to be considered are 10\% at 100 MeV, 5\% at 500 MeV and 20\% at 10 GeV (Rando et al. 2009).}
\tablenotetext{c}{Photon index of the power-law model used to fit LAT data.}
\tablenotetext{d}{Observed $\gamma-$ray luminosity.}
\tablenotetext{e}{Confidence level of the association according to the FoM.}
\tablenotetext{f}{See Abdo et al. (2009a).}
\normalsize
\end{table*}

In order to build SEDs, we retrieved all the publicly available data of these sources. In the case of PKS~1502+036, no X-ray data were found and, therefore, we asked for a $5$~ks snapshot with the \emph{Swift} satellite, which was performed on 2009 July 25 (ObsID 00031445001). We retrieved from the public \emph{Swift} archives, optical/UV/X-rays data also for PMN~J0948+0022 (Abdo et al. 2009a) and 1H~0323+342. For the latter, we analyzed all the observations performed between 2006 July 6 and 2008 November 16 (15 observations) and averaged the results (total exposure on XRT was 57~ks). In the case of PKS~2004-447, one \emph{XMM-Newton} observation was found (ObsID 0200360201, performed on 2004 April 11; 42~ks exposure on EPIC, see Gallo et al. 2006) and analyzed. 

\begin{table*}
\begin{center}
\scriptsize
\caption{Summary of the \emph{Swift} and \emph{XMM-Newton} analyses.\label{tab:swift}}
\begin{tabular}{lcccccccc}
\tableline\tableline
\multicolumn{9}{c}{X-rays}\\
Source & Inst\tablenotemark{a} & Exp\tablenotemark{b} & $N_{\rm H}$\tablenotemark{c} & $\Gamma$\tablenotemark{d} & $E_{\rm b}$\tablenotemark{e} & $\Gamma_2$\tablenotemark{f} & $F$\tablenotemark{g} & Stat\tablenotemark{h} \\
\tableline
1H 0323+342 & X & $57.1$ & $11.7$ & $2.03\pm 0.02$\tablenotemark{i} & -- & -- & $14.0\pm 0.1$ & $1.06/313$\\
PKS 1502+036 & X & $4.6$ & $3.89$ & $1.0_{-0.8}^{+0.9}$ &  -- & -- & $0.43\pm 0.10$	& $2$\\
PKS 2004-447 & E & $15,22,22$ & $2.96$ & $2.0_{-0.1}^{+0.2}$ & $0.67\pm 0.08$ & $1.50\pm 0.03$ & $1.5\pm 0.1$ & $1.05/315$\\
\tableline
\tableline
\multicolumn{9}{c}{Optical/UV}\\
Source & Inst\tablenotemark{a} & $A_V$\tablenotemark{l} & $v$ & $b$ & $u$ & $uvw1$ & $uvm2$ & $uvw2$\\
\tableline
1H 0323+342 & U & $0.62$ & $15.82\pm 0.03$ & $16.38\pm 0.03$ & $15.50\pm 0.03$ & $15.82\pm 0.04$ & $16.01\pm 0.04$ & $15.92\pm 0.04$\\
PKS 1502+036 & U & $0.21$ & $19.0\pm 0.2$ & $19.5\pm 0.1$ & $18.6\pm 0.1$ & $18.54\pm 0.08$ & $18.38\pm 0.08$ & $18.40\pm 0.06$\\
PKS 2004-447 & O & $0.16$ & -- & $18.91\pm 0.08$ & $18.54\pm 0.09$ & $18.65\pm 0.08$ & $19.0\pm 0.2$ & -- \\
\tableline
\tableline
\end{tabular}
\tablenotetext{a}{Instrument used for the observation: X for XRT and U for UVOT onboard \emph{Swift}; E for EPIC PN, MOS1, and MOS2 and O for OM onboard \emph{XMM-Newton}.}
\tablenotetext{b}{Net exposure in kiloseconds. In the case of EPIC, the exposures of PN, MOS1 and MOS2, respectively, are reported.}
\tablenotetext{c}{Galactic absorption in units of $10^{20}$~cm$^{-2}$ from Kablerla et al. (2005).}
\tablenotetext{d}{Photon index of the power-law model or low-energy photon index in the case of broken power-law model.}
\tablenotetext{e}{Break energy in the case of broken power-law model [keV].}
\tablenotetext{f}{High-energy photon index in the case of broken power-law model.}
\tablenotetext{g}{Observed flux in the $0.2-10$~keV energy band in units of $10^{-12}$~erg~cm$^{-2}$~s$^{-1}$.}
\tablenotetext{h}{Statistical parameters: $\tilde{\chi}^2/dof$ for $\chi^2$ or number of PHA bins for the Cash statistic.}
\tablenotetext{i}{It is worth noting that the addition of an unresolved gaussian emission line at $E=6.5\pm 0.3$ keV with equivalent width $\sim 147$~eV gives a $\Delta\chi^2=9.3$.}
\tablenotetext{l}{Magnitudes absorbed by the Galactic column in the visual filter (from Kalberla et al. 2005) and used as reference to calculate the absorption in other filters by means of the Cardelli et al. (1989) extinction law.}
\normalsize
\end{center}
\end{table*}

PMN~J0948+0022 was extensively studied in Abdo et al. (2009a,c), while the data of 1H~0323+342 and PKS~2004-447 were already discussed in Foschini et al. (2009b). Particularly, we found in the case of 1H~0323+342, there is evidence of spectral variability in the X-ray energy band, even on timescales of a few days. The photon index is generally soft, as for a typical NLS1 (cf Leighly 1999), but sometimes it displays a hard tail ($\Gamma \approx 1.3-1.6$) during a simultaneous increase of the optical/UV flux (Foschini et al. 2009b).

PKS~1502+036 was first observed in X-rays on 2009 July 25 (\emph{Swift} ObsID 00031445001). No previous X-ray observations were available and even \emph{ROSAT} measured just an upper limit, which was not very stringent ($<2\times 10^{-12}$~erg~cm$^{-2}$~s$^{-1}$ in the $0.1-2.4$~keV, see Yuan et al. 2008). We based our plans for the observation taking into account the similar optical characteristics shared with the better known PMN~J0948+0022 (Abdo et al. 2009a). However, the effective exposure of $4.6$~ks on XRT was not long enough to collect sufficient photons to apply the $\chi^2$ test statistic. Therefore, we grouped the photons to have at least 10 counts per bin and applied the Cash statistic, which allows the parameters to be estimated through the likelihood ratio (Cash 1979). 

In order to make a homogeneous data set, we have reanalyzed all the \emph{Swift} and \emph{XMM-Newton} data adopting the same procedures outlined in the previous works, but using the most recent software packages version (\texttt{HEASOFT v. 6.6.3} for \emph{Swift} and \texttt{SAS v. 9.0.0} for \emph{XMM-Newton}) and calibration databases (updated on 2009 June 5 for \emph{Swift} and 2009 June 11 for \emph{XMM-Newton}). X-ray data were fitted to power-law or broken power-law models with Galactic absorption from Kalberla et al. (2005), while optical/UV data were dereddened for the proper values calculated according to the extinction laws by Cardelli et al. (1989). None of the sources showed evidence of X-ray absorption in excess to the Galactic value. The X-ray photon indexes are generally in agreement with those typical of blazars (c.f., for example, Foschini et al. 2006; Maraschi et al. 2008). The results are summarized in Table~\ref{tab:swift}. All the results of the present reanalysis are consistent with the previous ones already reported in Foschini et al. (2009b).

\begin{figure*}[!ht]
\centering
 \begin{minipage}{0.4\textwidth}
   \centering
   \includegraphics[scale=0.4,clip,trim = 20 50 10 50]{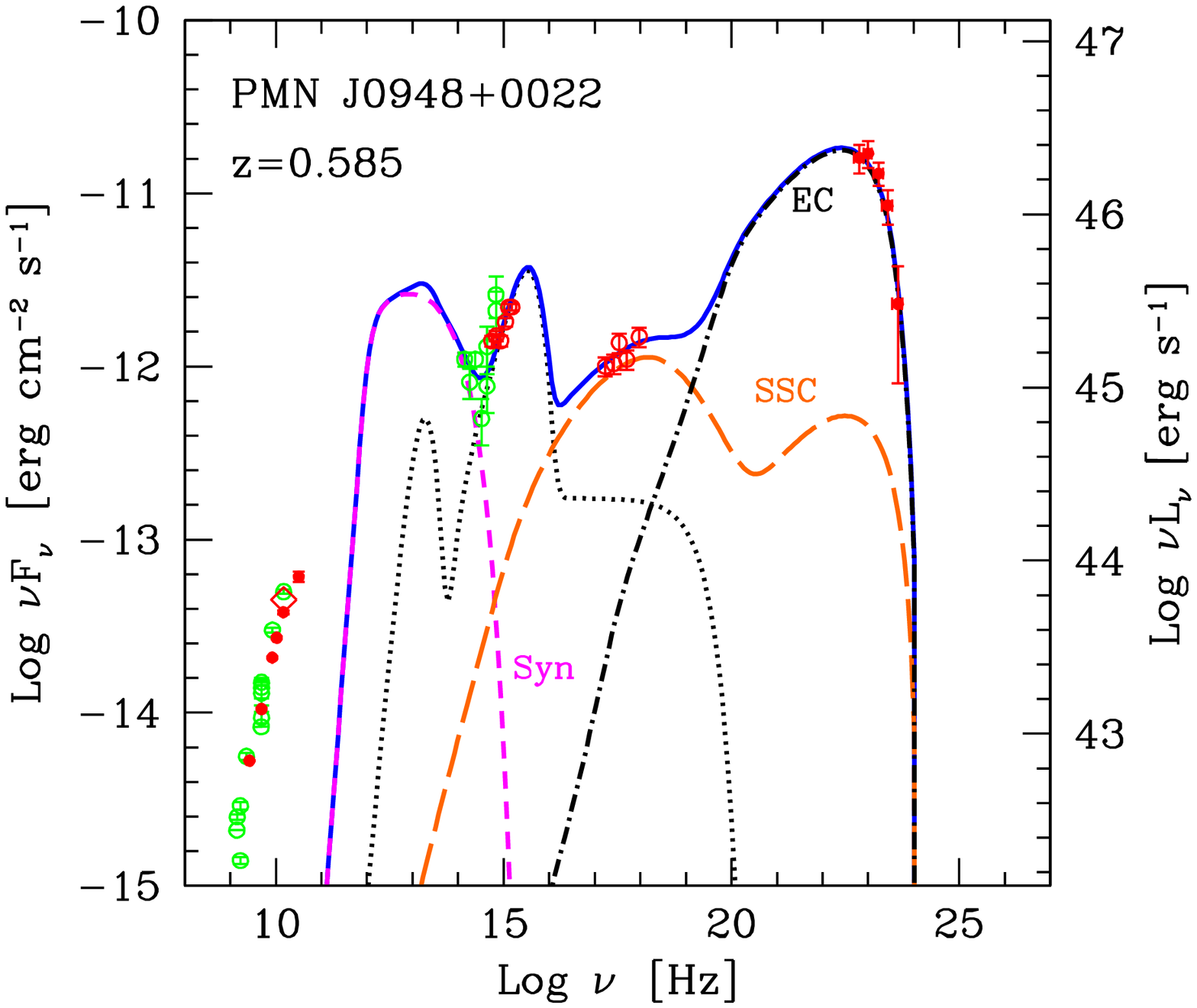}
 \end{minipage}
 \ \hspace{2mm} \hspace{3mm} \
 \begin{minipage}{0.4\textwidth}
  \centering
   \includegraphics[scale=0.4,clip,trim = 20 50 10 50]{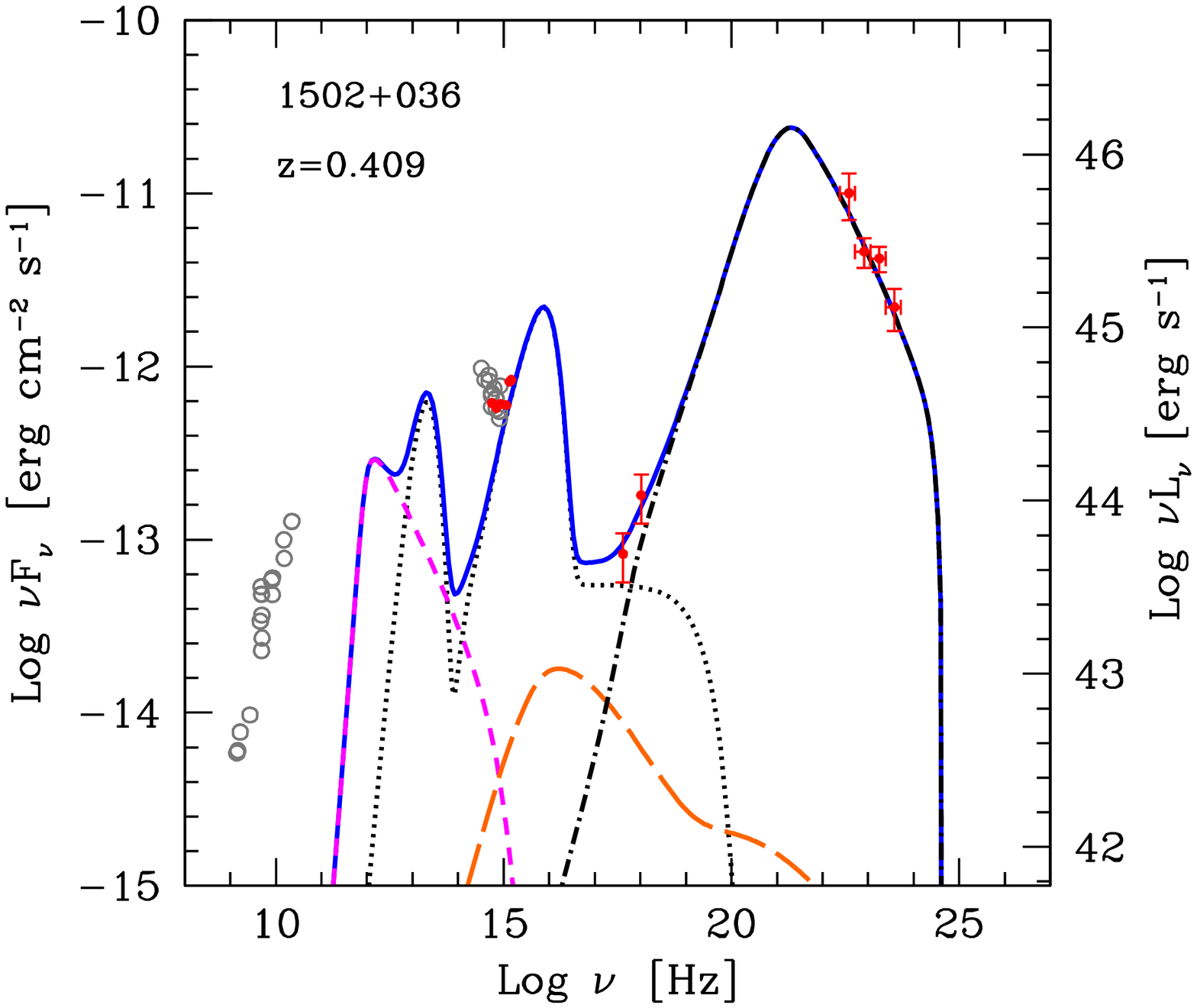}
 \end{minipage}\\
 
  \begin{minipage}{0.4\textwidth}
   \centering
  \includegraphics[scale=0.4,clip,trim = 20 50 10 50]{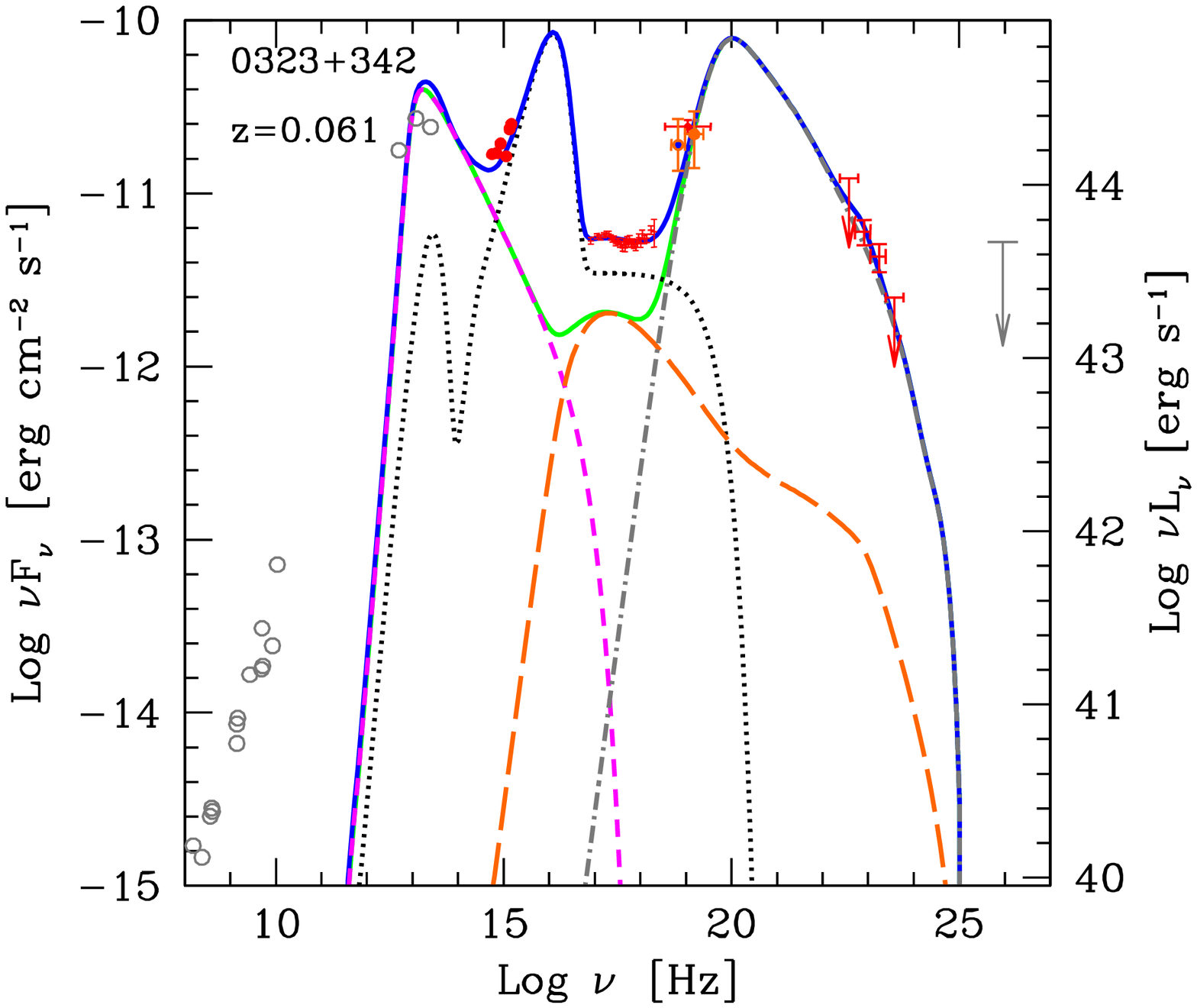}
 \end{minipage}
 \ \hspace{2mm} \hspace{3mm} \
 \begin{minipage}{0.4\textwidth}
  \centering
   \includegraphics[scale=0.4,clip,trim = 20 50 10 50]{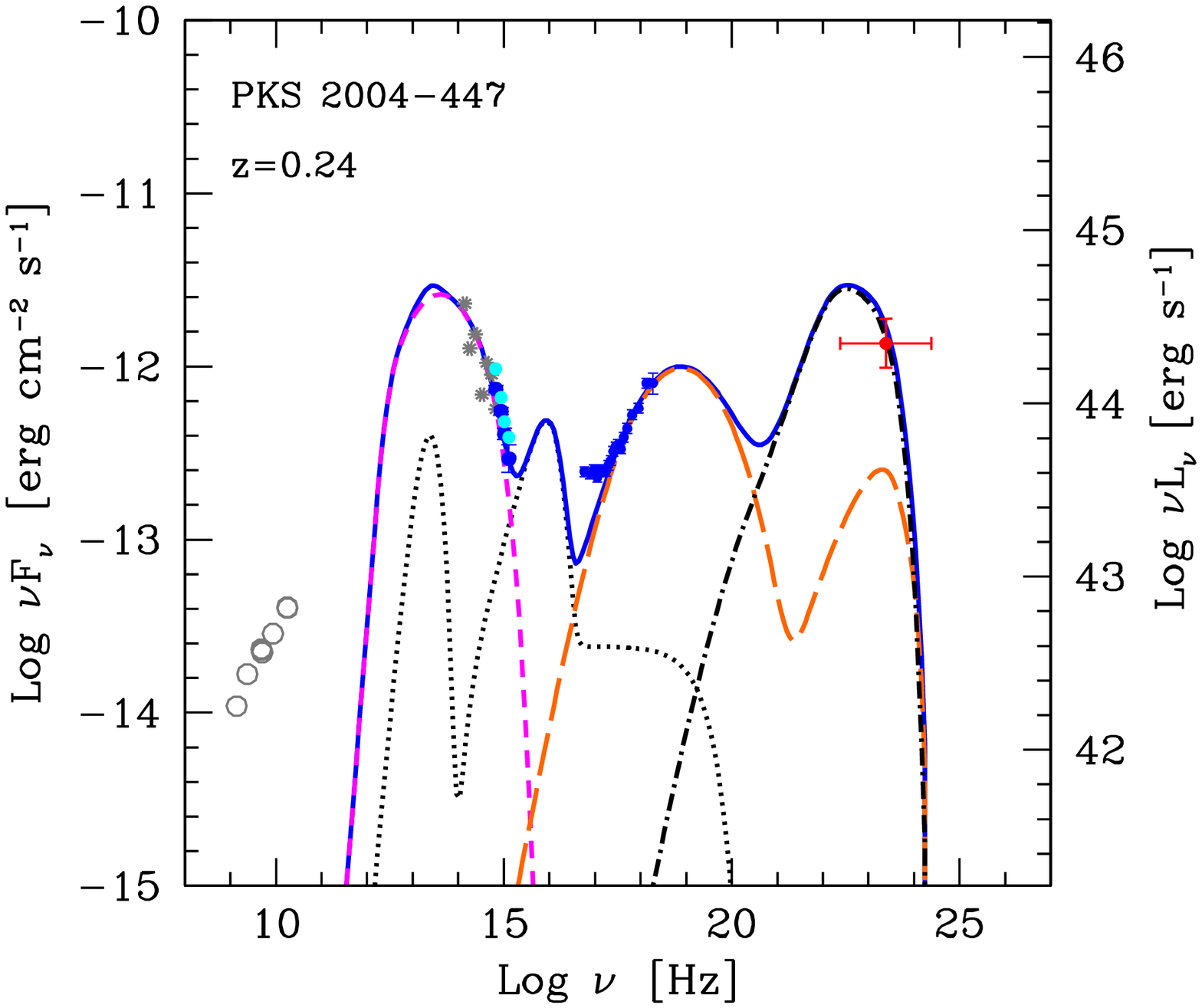}
 \end{minipage}
\caption{SEDs of the four RL-NLS1s detected by \emph{Fermi}/LAT in order of decreasing $\gamma-$ray flux. The SED of PMN J0948+0022 is that from Abdo et al. (2009a). In the SED of 1H 0323+342, there are also detections in the hard X-rays by \emph{Swift}/BAT (Cusumano et al. 2009) and \emph{INTEGRAL}/ISGRI (Bird et al. 2007). The TeV upper limit is derived from observations with \emph{Whipple} (Falcone et al. 2004). Archival radio data are from NED, but are displayed just for completeness, since they are not used in the model fit. The synchrotron self-absorption is clearly visible around $10^{11-12}$~Hz. The short dashed light blue line indicates the synchrotron component, while the long dashed orange line is the SSC. The dot-dashed line refers to EC and the dotted black line represents the contribution of the accretion disk, X-ray corona and the IR torus. The continuous blue line is the sum of all the contributions. \label{fig:SED}}
\end{figure*}

\begin{table*}[!h]
\caption{Parameters used to model the SEDs. \label{tab:inputmodel}}
\vskip 0.4 true cm
\scriptsize
\centering
\begin{tabular}{lcccccccccccc}
\hline
\hline
Name  &$R_{\rm diss}$\tablenotemark{a} &$\log M$\tablenotemark{b} &$R_{\rm BLR}$\tablenotemark{c} &$P^\prime_{\rm i}$\tablenotemark{d} &$L_{\rm d}$\tablenotemark{e} &$B$\tablenotemark{f} &$\Gamma_{\rm b}$\tablenotemark{g} &$\theta_{\rm v}$\tablenotemark{h} &$\gamma_{\rm e,break}$\tablenotemark{i} &$\gamma_{\rm e,max}$\tablenotemark{l} &$s_1$\tablenotemark{m}  &$s_2$\tablenotemark{n}  \\
\hline   
1H 0323+342    &1.9 (650)  & 7.0   &116  & 1.0    &1.4 (0.9) &30 &12 &3  &60  &6000   &--1 &3.1  \\
PKS 1502+036    &24 (4000)   & 7.3   &155  & 21     &2.4  (0.8) &1.6 &13 &3  &50  &3000   &1   &3.2  \\
PKS 2004--447   &6  (4000)   & 6.7\tablenotemark{o}   &39   & 3.1    &0.15 (0.2) &6.9  &8  &3  &120 &1500 &0.2 &2    \\
\hline
PMN J0948+0022\tablenotemark{*}   &72 (1600) & 8.2 &300  &  240       &9  (0.4)    &3.4  &10 &6  &800 & 1600 &1  &2.2    \\
\hline
\hline 
\end{tabular}
\tablenotetext{a}{Dissipation radius in units of $10^{15}$~cm and (in parenthesis) in units of the Schwarzschild radius.}
\tablenotetext{b}{Black hole mass in units of $M_{\odot}$ (details and caveats about the mass estimation used in this work can be found in Ghisellini et al. 2009a; the error with this method is generally about 50\%).}
\tablenotetext{c}{Size of the BLR in units of $10^{15}$~cm.}
\tablenotetext{d}{Power injected in the blob calculated in the comoving frame, in units of $10^{41}$~erg~s$^{-1}$.}
\tablenotetext{e}{Accretion disk luminosity in units of $10^{45}$~erg~s$^{-1}$ calculated by integrating the thermal component (black dotted line) of the SEDs in Fig.~\ref{fig:SED} and (in parenthesis) in Eddington units.}
\tablenotetext{f}{Magnetic field in Gauss.}
\tablenotetext{g}{Bulk Lorentz factor at $R_{\rm diss}$.}
\tablenotetext{h}{Viewing angle in degrees.}
\tablenotetext{i}{Break random Lorentz factors of the injected electrons.}
\tablenotetext{l}{Maximum random Lorentz factors of the injected electrons.}
\tablenotetext{m}{Slope of the injected electron distribution below $\gamma_{\rm e,break}$.}
\tablenotetext{n}{Slope of the injected electron distribution above $\gamma_{\rm e,break}$.}
\tablenotetext{o}{Fixed from measurement with reverberation mapping method reported by Oshlack et al. (2001).}
\tablenotetext{*}{See Abdo et al. (2009a).}
\normalsize
\end{table*}

\begin{table}[!h]
\caption{Power carried out by the jet.\label{tab:outputmodel}} 
\vskip 0.4 true cm
\centering
\scriptsize
\begin{tabular}{lcccc}
\hline
\hline
Name   &$\log P_{\rm r}$\tablenotemark{a} &$\log P_{\rm B}$\tablenotemark{b} &$\log P_{\rm e}$\tablenotemark{c} &$\log P_{\rm p}$\tablenotemark{d} \\
\hline   
1H 0323+342  &42.8 &43.3 &42.7 &44.3 \\  
PKS 1502+036  &44.0 &43.0 &44.1 &46.2  \\ 
PKS 2004+447  &42.9 &42.6 &42.9 &44.1 \\ 
\hline  
PMN J0948+0022\tablenotemark{*} &45.3 &44.3 &44.7 &46.7 \\    
\hline
\hline 
\end{tabular}
\tablenotetext{a}{Radiative power [erg~s$^{-1}$].}
\tablenotetext{b}{Poynting flux power [erg~s$^{-1}$].}
\tablenotetext{c}{Power in bulk motion of electrons [erg~s$^{-1}$].}
\tablenotetext{d}{Power in bulk motion of protons, assuming one proton per emitting electron [erg~s$^{-1}$].}
\tablenotetext{*}{See Abdo et al. (2009a).}
\normalsize
\end{table}

\section{Spectral Energy Distributions (SED)}
The SEDs built with all the available data are displayed in Fig.~\ref{fig:SED}. As for PMN J0948+0022, these SEDs show clear similarities with blazars and therefore we fitted them with the synchrotron and inverse-Compton model developed by Ghisellini \& Tavecchio (2009). Basically, it adopts an injected relativistic electron distribution with a broken power-law shape, of parameters $\gamma^{-s_1}_{\rm e}$ for energies below the break $\gamma_{\rm e,break}$ and $\gamma^{-s_2}_{\rm e}$ above, where $\gamma_{\rm e}$ is the random Lorentz factor of electrons. This distribution is iteratively modified to take into account electron cooling and the possibility to produce pairs through $\gamma\gamma \rightarrow e^{\pm}$. The resulting distribution at the time $R/c\approx 0.1R_{\rm diss}/c$, where $R_{\rm diss}$ is the dissipation radius and $0.1R_{\rm diss}$ is the size of the emitting spherical blob, is used to generate the radiation emitted through the processes of synchrotron, synchrotron self-Compton (SSC) and external-Compton (EC). The seed photons for the latter are the sum of several contributions: photons directly radiated from the accretion disk (Dermer \& Schlickeiser 1993), from the broad-line region (BLR\footnote{In the case of NLS1s, the permitted emission lines from the BLR are narrower than usual, with FWHM(H$\beta$)$~\lesssim 2000$~km~s$^{-1}$.}, Sikora, Begelman \& Rees 1994) and from the infrared torus (B{\l}a\.zejowski et al. 2000; Sikora et al. 2002). We refer to Ghisellini \& Tavecchio (2009) for more details. The model parameters and the calculated powers are listed in Tables~\ref{tab:inputmodel} and \ref{tab:outputmodel}. 

Archival radio data are displayed in Fig.~\ref{fig:SED} for completeness, but are not used in the model fit. Indeed, the one-zone model used in the present work has to explain the bulk of the emission and, therefore, necessarily requires a compact source. This, in turn, is self-absorbed for synchrotron radiation (at $\approx 10^{11-12}$~Hz). The radio emission is detectable only when the blob becomes optically thin and this occurs as it moves and expands, further out in the jet (e.g. Blandford \& K\"onigl, 1979).

Given the scarce, non-simultaneous data and the weakness of the $\gamma-$ray emission for the three newly discovered sources, it is not possible to tightly constrain the selection of parameters. As a guide to the selection of the initial values of the model parameters (Table~\ref{tab:inputmodel}), we were driven by the recent (March-July 2009) multiwavelength study of PMN J0948+0022 (Abdo et al., 2009c), our knowledge of blazars, and the mass measurements performed with other methods (particularly in the case of PKS 2004-447, for which the optical/UV data are not sampling the accretion disk emission). When more information, particularly from multiwavelength variability studies, will be available, it will be possible to improve the estimation of the parameter values.

\section{Discussion}
The analysis of the individual sources reveals some things worth noting. In 1H 0323+342 the dissipation region needs to be located very close to the central black hole ($\sim 650$ times the Schwarzchild radius $R_{\rm S}$) and hence a quite high magnetic field (up to $30$ gauss) is needed. We note that PKS 2004-447 seems to be different with respect to the three other RL-NLS1s, with optical/UV data sampling the synchrotron emission instead of the accretion disk. Indeed, the classification of this source is still open: while Oshlack et al. (2001) classified it as genuine RL-NLS1, Komossa et al. (2006) suggested it can be a narrow-line radio galaxy, on the basis that the bump of FeII is weak. Gallo et al. (2006) noted that there is no specific prescription on the value of the strength of the FeII bump and confirmed the NLS1 classification. The same authors noted that PKS 2004-447 is a compact steep-spectrum (CSS) radio source. On the other hand, this source is also in the CRATES catalog of flat spectrum radio sources (Healey et al. 2007), which includes also the three others NLS1s in this work. From the present study, the estimated jet power of PKS 2004-447 is well above the range typical of radio galaxies and, therefore, we favor the hypothesis of a genuine NLS1. However, in this case the classification remains uncertain and should be studied with further observations. 

A synoptic analysis of the 4 RL-NLS1s in the present work shows that PKS 1502+036 carries a jet power comparable to PMN J0948+0022, while the remaining two sources are significantly less powerful -- even two order of magnitudes for proton powers (Table~\ref{tab:outputmodel}). The comparison with the large sample of blazars reported in Celotti \& Ghisellini (2008) and Ghisellini et al. (2009b), shows that the jet powers of PKS 1502+036 and PMN J0948+0022 are in the region of quasars, while 1H 0323+342 and PKS 2004-447 are in the range typical of BL Lac Objects. The $\gamma-$ray emitting RL-NLS1s in our sample have small masses and high accretion rates. This is in contradiction to the larger masses and and smaller accretion rates expected from low-power blazars (HFSRQs) with strong emission lines, as suggested by Yuan et al. (2008). Anyway, when comparing the calculated powers with the distribution of powers in blazars (quasars plus BL Lacs; see Ghisellini et al. 2009b), RL-NLS1s are in the average range. 

The main differences with respect to blazars are in the masses and accretion rates. The masses of the three newly discovered RL-NLS1s are around $10^{7}M_{\odot}$, in agreement within one order of magnitude with the values obtained with other methods. In the case of 1H 0323+342, Zhou et al. (2007) found values of $1.8\times 10^{7}M_{\odot}$ with $H\beta$ luminosity and $3\times 10^{7}M_{\odot}$ with the luminosity of the continuum at 5100\AA. In the case of PKS 1502+036, Yuan et al. (2008) estimate the mass to be $4\times 10^{6}M_{\odot}$, by means of the virial method. These value are about 1-2 orders of magnitude lower than the typical blazar masses ($\approx 10^{9}M_{\odot}$, see Ghisellini et al. 2009b). The accretion rates can reach extreme values, up to 80 or even 90\% the Eddington values in the cases of PKS 1502+036 and 1H 0323+342, respectively. These values are the most extreme ever found in any $\gamma-$ray emitting AGNs, but usual for NLS1s. 

What is at odds with this scenario is the type of host galaxy, which is elliptical in all blazars, while is likely to be spiral in RL-NLS1s (see, e.g. Zhou et al. 2006). In the case of 1H 0323+342, analysis of optical observations suggests two possibilities: Zhou et al. (2007) show the spiral arms of the host galaxy by means of observations with the \emph{Hubble Space Telescope}, while Ant\`on et al. (2008), on the basis of ground-based observations with Nordic Optical Telescope (NOT), suggest that these structures are the residual of a merging that occurred within the past $10^8$ years. This means that relativistic jets can form and develop independently of their host galaxies, with quasars, BL Lacs and, now, RL-NLS1s jointly characterized by the presence of a relativistic jet and with the differences in their observed SEDs mainly determined by their masses and accretion rates.

\acknowledgments
The \textit{Fermi} LAT Collaboration acknowledges generous ongoing support
from a number of agencies and institutes that have supported both the
development and the operation of the LAT as well as scientific data analysis.
These include the National Aeronautics and Space Administration and the
Department of Energy in the United States, the Commissariat \`a l'Energie Atomique
and the Centre National de la Recherche Scientifique / Institut National de Physique
Nucl\'eaire et de Physique des Particules in France, the Agenzia Spaziale Italiana
and the Istituto Nazionale di Fisica Nucleare in Italy, the Ministry of Education,
Culture, Sports, Science and Technology (MEXT), High Energy Accelerator Research
Organization (KEK) and Japan Aerospace Exploration Agency (JAXA) in Japan, and
the K.~A.~Wallenberg Foundation, the Swedish Research Council and the
Swedish National Space Board in Sweden. Additional support for science analysis during the operations phase is gratefully
acknowledged from the Istituto Nazionale di Astrofisica in Italy and the Centre National d'\'Etudes Spatiales in France.

This research has made use of the NASA/IPAC Extragalactic Database (NED) which is operated by the Jet Propulsion Laboratory, California Institute of Technology, under contract with the National Aeronautics and Space Administration and of data obtained from the High Energy Astrophysics Science Archive Research Center (HEASARC), provided by NASA's Goddard Space Flight Center.

\end{document}